\documentclass[a4paper,showkeys,floatfix,aps,pre,reprint,groupedaddress]{revtex4-1}
\usepackage{graphicx}
\usepackage{amsmath, amssymb, amsfonts}
\usepackage{color}
\usepackage{subfig}
\usepackage{caption}
\usepackage{comment}

\begin{document}

\title{Dynamical percolation transition in the Ising model studied using a pulsed magnetic field}



\author{Soumyajyoti Biswas}
\email[]{soumyajyoti.biswas@saha.ac.in}
\author{Anasuya Kundu}
\email[]{anasuya.kundu@saha.ac.in}
\author{Anjan Kumar Chandra}
\email[]{anjan.chandra@saha.ac.in}

\affiliation{
Theoretical Condensed Matter Physics Division, Saha Institute of Nuclear 
Physics, 1/AF Bidhannagar, Kolkata-700064, India.\\
}

\date{\today}

\begin{abstract}
\noindent We study the dynamical percolation transition of the geometrical clusters in the 2d Ising model when it is subjected to a pulsed field below the critical temperature. The critical exponents are independent of the temperature and pulse width and are different from the (static) percolation transition associated with the thermal transition. For a different model that belongs to the Ising universality class, the exponents are found to be same, confirming that the behavior is a common feature of the Ising class. These observations, along with a universal critical Binder cumulant value, characterises the dynamical percolation of the Ising universality class. 
\end{abstract}


\maketitle

\section{Introduction}
\noindent The thermal phase transition of the Ising model have been extensively studied in the geometrical terms of (correlated) percolation in the past. It was seen that the ``geometrical'' clusters, consisting of nearest neighbor parallel spins, undergo a percolation transition \cite{muler1,stoll,muler2,binder1} at the same critical temperature as the magnetisation in two dimensions \cite{coniglio1, coniglio2}. In higher dimensions, however, the transition point differs \cite{muler1, heermann}. Even for the two dimensional case, although the critical points coincide, the critical behavior does not \cite{sykes, stella}. Later, by redefining the clusters  suitably \cite{coniglio2} (the so called ``physical'' clusters), the critical point was made to coincide in all dimensions along with the critical behavior \cite{hu}. 

Although the temperature induced percolation transition is well studied, very little is known about the percolation transition induced by external magnetic field. It is well known \cite{chat, bkc1, bkc2, bkc3, bkc4, bkc5, bkc6} that pure Ising systems, below its static critical temperature ($T_c^0$) can undergo a field-pulse induced magnetisation reversal transition. The field pulse is applied to the Ising system in the direction opposite to that of the existing order for a finite time ($\Delta t$). Depending on the value of the field amplitude ($h_p$), temperature ($T$) and pulse duration ($\Delta t$), the system can eventually undergo a transition from one equilibrium state (with magnetisation $-m_0$) to the other (with magnetisation $m_0$) after the field is withdrawn. Apart from its theoretical interests, it has experimental realisations \cite{expt} as well as industrial applications, such as in recording and switching industry (for  review see \cite{bkc7, bkc8}). 

As there is a percolation transition associated with the thermal one, here also we expect a similar percolation transition. In the following study, we investigate this percolation transition of the ``geometrical'' clusters at different temperatures, field pulse amplitude and field pulse widths. We  find  the critical exponents associated with the percolation transition to be significantly different from the percolation transition of the ``geometrical'' clusters in the thermal (or static) transition. These exponents remain invariant with change in temperature and field pulse width. We have also measured the critical Binder cumulant value \cite{binder2} of the percolation transition order parameter. This quantity is not only useful in locating the transition point but also its value at the critical point (which is independent of system size) indicates the universality class to which the transition belongs. It is seen that this value is different from the thermal counterpart and is very robust in the sense that it remains unchanged with changes in temperature and field pulse width. 

It was shown before \cite{fort2,fort} for the static transition that the percolation exponents of the geometrical clusters have same values for the models belonging to the same universality class (the case of Ising and $Z(3)$ symmetric models were studied). For all models belonging to Ising class, therefore, universal exponents were found. In our study of dynamical percolation transition also, if the exponents have to characterise the dynamical percolation transition of the Ising universality class, their values have to be same for all models belonging to this class. We have checked for a different model (in Ising class) that the exponents and  the critical Binder cumulant value are the same.      

For the magnetisation reversal transition also, it was claimed \cite{chat, bkc8} that the universality class was different from the static one. In that case, although the exponents were robust with respect to changes in temperature and pulse width, we find that the critical Binder cumulant value changes continuously with temperature, leaving the claim  incomplete. However, as the associated percolation transition (occurring at the same critical point within our numerical accuracy) clearly belongs to a different universality class, one can probably say that the claim of different universality class there was actually true. And the non-universal critical Binder cumulant value can be put as yet another example of some special cases, which is known to occur also for changes in boundary condition, lattice shape, anisotropic effect etc. \cite{binder4, kam, janke, chen, selke1, selke2, selke3, dohm, selke4}.

The rest of the paper is organised as follows, in sec. II we define the model and the percolation transition of the ``geometrical'' clusters. In sec. III we measure the critical exponent and Binder cumulant values for different temperatures and pulse widths and compare them with the magnetisation reversal transition. In sec. IV we discuss our main results and conclude.

\section{Field pulse induced percolation transition}

\subsection {Pure Ising model}

\noindent Here we have studied the percolation transition induced by an external magnetic field pulse in the case of pure two dimensional Ising model with nearest neighbour interaction. The Hamiltonian of the system reads
\begin{equation}
\label{dpe1}
H=-\sum _{\{ij\}}J_{ij}S_{i}S_{j}-h(t)\sum _{i}S_{i},
\end{equation}

\noindent where $S_i$'s ($i=1,2,\dots,N$) are Ising spins on lattice site $i$ and $J_{ij}$ is the cooperative interaction between the nearest neighbor spins. Since we have considered a square lattice, the static critical point $T_c^0=2.269...$ (in units of $J/k_B$). Periodic boundary condition is used along both directions. We keep the temperature $T$ of the system below this critical value such that there is a spontaneous equilibrium magnetisation. The time dependent external magnetic field $h(t)$ is applied for a finite duration in the direction opposite to that of the equilibrium magnetisation. Although spatially uniform, the time dependence of this applied field is as follows:
\begin{equation}
h(t)=\left\{ \begin{array}{cc}
h_{p}, & t_{0}\leq t\leq t_{0}+\Delta t\\
0, & \rm otherwise.
\end{array}\right. 
\end{equation}

\begin{figure}[tbh]
\begin{center}
 \includegraphics[width=0.7\linewidth, angle=270]{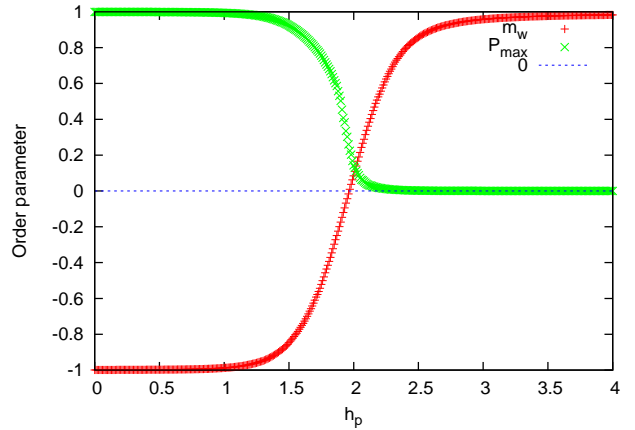}
\end{center}
   \caption{(Color online) Variations of the order parameters in the two transitions (a) withdrawal point magnetisation ($m_w$) and (b) the probability of largest cluster ($P_{max}$) with the field pulse amplitude ($h_p$) in the case of pure Ising model, where $L=100, T=1.0$ and $\Delta t=4$.}
\label{ordp}
\end{figure}

\noindent In a given Monte Carlo (MC) update a spin is randomly selected and the energy difference ($\Delta E$) caused for flipping it is calculated. The flip is made if a random number is less than $\exp(-\Delta E/T)$. $L^2$ such updates make a single MC step. Here we take the geometrical definition of the clusters where the clusters are formed from parallel spins and each spin in a cluster must have at least one of its neighbors parallel to it. At $t=t_0$, there is a spontaneous magnetisation (say, in the ``downward'' direction) in the system and one spanning cluster (of ``down'' spins) is present. Then magnetic field is applied in the opposite direction (``upward'') for a finite duration ($\Delta t$) and the absolute value of the magnetisation as well as the percolation order parameter $P_{max}=S_L/L^2$  (where $S_L$ is the size of the largest cluster and $L$ is the linear size of the system) decreases with time in that duration. For a particular combination of $T$ and $\Delta t$ at some $h_p^c(T,\Delta t)$, the system undergoes a percolation transition as more and more spins flip upward and the previously percolating cluster (of down spins) becomes non-percolating (see Fig.~\ref{ordp}). In the following, we have studied the critical behavior of this percolation transition at different points in the $h_p-\Delta t$ phase boundary (details are shown for one point only) which coincides with (within our numerical accuracy) the phase boundary obtained before \cite{chat} in the magnetisation reversal transition (where the order parameter was defined \cite{chat} as the magnetisation  $m_w$ at which the field is withdrawn).

\subsection{Ising model with diagonal second neighbor frustration}
\noindent To verify if the transition mentioned above characterise the dynamical percolation of the Ising universality class, here we study the dynamical percolation in a different model that belongs to the Ising universality class. This model of Ising spins has nearest neighbor (NN) ferromagnetic and (diagonal) next nearest neighbor (NNN) (weaker) antiferromagnetic coupling

\begin{equation}
H=-J_1\sum\limits_{NN}S_iS_j+J_2\sum\limits_{NNN}S_iS_j+h(t)\sum\limits_iS_i.
\end{equation}
We study this model for $J_2/J_1=1/10$ (as was chosen in \cite{fort2}) and $h(t)$ following Eqn. (2). 
The static (when $h(t)=0$) critical temperature for the order-disorder transition in this model is $T_c \approx 1.945$ \cite{fort2} in the same unit as before. We perform the following study below this critical temperature (at $T=1.0$). At that temperature, the system is in ferromagnetic order. As before, a magnetic field pulse is applied in the direction opposite to this order for a finite duration. For a fixed value of temperature and pulse width (4 MC time steps), by increasing the field pulse amplitude $h_p$, the previously spanning cluster of parallel spins becomes non-spanning and also the system undergoes a magnetisation reversal transition (Fig 2).   

\begin{figure}[tbh]
\begin{center}
 \includegraphics[width=0.7\linewidth, angle=270]{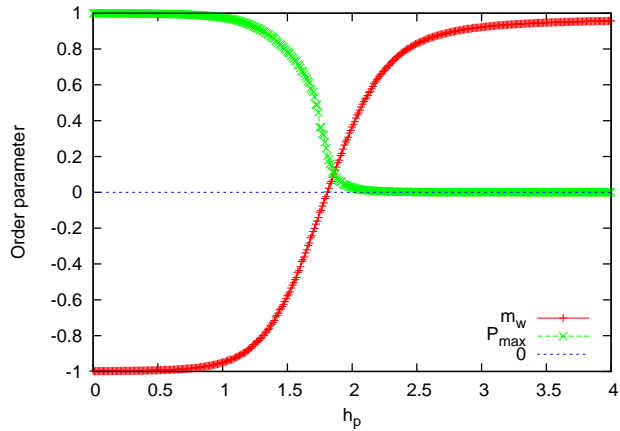}
\end{center}
   \caption{(Color online) Variations of the order parameters of magnetisation reversal ($m_w$) and the percolation transition ($P_{max}$) with the field pulse amplitude ($h_p$), where $L=100, T=1.0$ and $\Delta t=4$ in the case of Ising model with next nearest neighbor frustration.}
\label{new-ordp}
\end{figure}

It should be mentioned here that when there is no external magnetic field, the thermal transition and the percolation transition of the geometrical clusters can be shown \cite{coniglio1, coniglio2} to occur at the same critical point in two dimensions.  In our case, however, the transition is driven by field which is time dependent. Hence, the evidence of coincidence in this case is entirely numerical.

\section{Critical behavior: finite size scaling}
\noindent The percolation transition is characterised by power-law variation of different quantities. The order parameter i.e., the relative size ($P_{max}$) of the largest cluster varies as 
\begin{equation}
P_{max}\sim (h_p^c-h_p)^{\beta}.
\end{equation}

\noindent The correlation length diverges near the percolation transition point as
\begin{equation}
\xi \sim (h_p^c-h_p)^{-\nu},
\end{equation}
\noindent where, $h_p^c$ is the critical field amplitude.

\noindent The values of the critical exponents $\beta$ and $\nu$  specify the universality class of the transition. The other exponents can be obtained from  scaling relations \cite{sa}.

\begin{figure}[tbh]
\begin{center}
 \centering
 \includegraphics[width=0.9\linewidth]{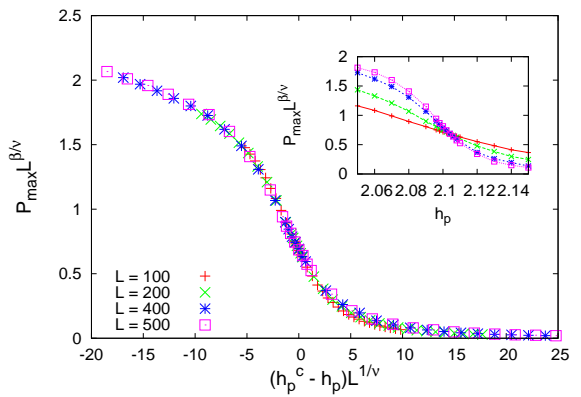}
\end{center}
   \caption{(Color online) Inset: $P_{max}L^{\beta/\nu}$ is plotted against the field amplitude ($h_p$) where $T=1.0$ and $\Delta t=4$ for pure Ising model. The curves for different system sizes ($L=100, 200, 400, 500$) cross at $h_p^c=2.105\pm0.005$ giving the critical point and $\beta/\nu=0.20\pm0.05$. Front: The value obtained for $\beta/\nu$ is used in the y-axis and by tuning the value of $1/\nu$ all the curves for different system sizes were made to collapse on a single curve, thereby estimating $1/\nu=0.85\pm0.05$. Error bars are smaller than the symbol size.}
\label{betanu}
\end{figure}


However, the exponents are not determined from these definitions due to finite size effects. The critical exponents are determined from the finite size scaling relations. For example, the order parameter is expected to follow the scaling form
\begin{equation}
P_{max}=L^{-\beta/\nu}\mathcal{F}\left[L^{1/\nu}\left (h_p^c-h_p\right)\right],
\end{equation}   

\noindent where $\mathcal{F}$ is a suitable scaling function. If we plot $P_{max}L^{\beta/\nu}$ against $h_p$ for different system sizes, then by tuning $\beta/\nu$, all the curves can be made to cross at a single point. The field amplitude where this happens must be the critical field amplitude ($h_p^c$). To estimate $\nu$,  $P_{max}L^{\beta/\nu}$ is to be plotted against $(h_p^c-h_p)L^{1/\nu}$ and by tuning $1/\nu$, the curves are made to collapse, giving an accurate estimate of the exponent $\nu$. 

To further verify the critical point and the universality class, we study the reduced fourth order Binder cumulant of the order parameter, defined as \cite{binder2}
\begin{equation}
U=1-\frac{\langle P_{max}^4\rangle}{3\langle P_{max}^2\rangle^2},
\end{equation}   

\noindent where $P_{max}$ is the percolation order parameter (as defined before) and the angular brackets denote ensemble average. $U \to \frac{2}{3}$ deep inside the ordered phase and $U\to 0$ in the disordered phase when the fluctuation is Gaussian.

The Binder cumulant can also be used to find the correlation length exponent as it follows the scaling form
\begin{equation}
\label{binder-nu}
U=\mathcal{U}((h_p^c-h_p)L^{1/\nu}),
\end{equation}

\noindent where $\mathcal{U}$ is a suitable scaling function.
\subsection{Pure Ising model}
\noindent In Fig.~\ref{betanu} (inset) we  plot $P_{max}L^{\beta/\nu}$ against $h_p$ for a given temperature ($T=1.0$) and field pulse width ($\Delta t=4$), where the unit of time is measured by Monte Carlo Steps (one MCS being $L^2$ spin updates). The curves for different system sizes cross at a given point when $\beta/\nu=0.20\pm 0.05$ and the crossing point ($h_p^c=2.105\pm0.005$) gives the critical field amplitude. 

To estimate the exponent $\nu$, we plot for the same set, $P_{max}L^{\beta/\nu}$ against $(h_p^c-h_p)L^{1/\nu}$ (see Fig.~\ref{betanu}). Now by tuning the $1/\nu$ term in the $X$-axis only, all the plots are made to collapse on a single curve. This gives us the estimate of $1/\nu=0.85\pm 0.05$. Again, the error bar is estimated by changing the tuning parameter upto the point when the scaling is visibly worsened. 
These estimates differ significantly from the corresponding static transition values reported in Ref.~\cite{fort2} ($\beta_s/\nu_s=0.052\pm 0.002$,
 $\nu_s=1.004\pm 0.009$).
\begin{figure}[tbh]
\begin{center}
 \centering
 \includegraphics[width=0.7\linewidth, angle=270]{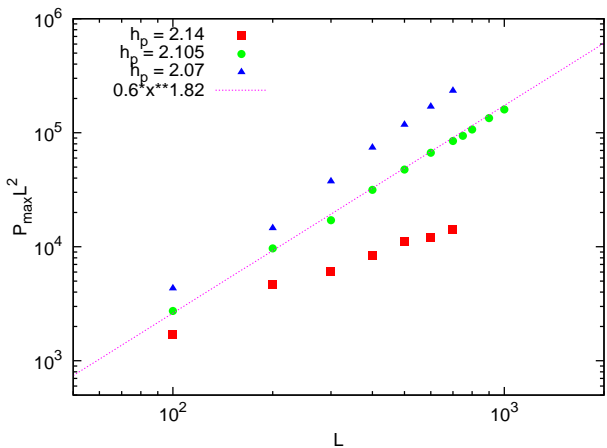}
\end{center}
   \caption{(Color online) Size of the largest cluster for various system sizes ($L=100, 200, 300, 400, 500, 600, 700, 800, 900, 1000$) at the critical field ($h_p^c=2.105$) where $T=1.0$ and $\Delta t=4$ for pure Ising model. The slope of the log-log plot gives the fractal dimension to be $D=1.82\pm0.01$. The other two curves show the variation of the same, below ($h_p=2.07$) and above ($h_p=2.14$) the critical field.}
\label{fractal}
\end{figure}


It is  known \cite{sa}  that the largest cluster at critical point has a fractal nature. One can find its fractal dimension from the relation $S_L\sim L^D$, where $S_L$ is the size of the largest cluster and $D$ is its fractal dimension. The fractal dimension is related to the spatial dimension of the lattice through the critical exponents in the following way $D=d-\beta/\nu$, where $d$ is the spatial dimension. While for static transition in Ising model the fractal dimension of the critical droplet is $1.875$ (for geometric cluster in static transition it is $1.947\pm0.002$ \cite{fort2}), from the best fit (Fig.~\ref{fractal}) of the largest cluster size with linear size in a log-log plot  gives the value to be $1.82\pm0.02$ in our case, which also agrees well with our estimate of $\beta/\nu$ from finite size scaling ($0.18$ and $0.20$ respectively).  

\begin{figure}[tbh]
\begin{center}
 \centering
 \includegraphics[width=0.9\linewidth]{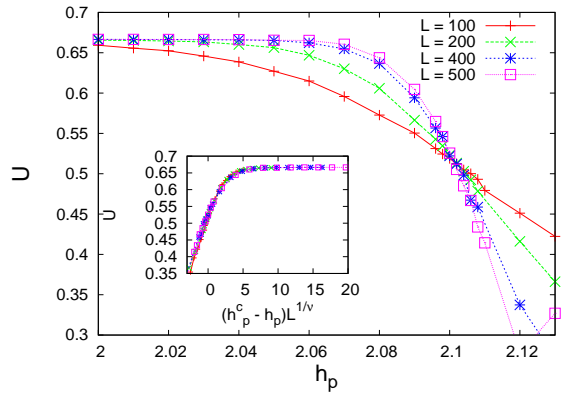}
\end{center}
   \caption{(Color online) Fourth order reduced Binder cumulant of the percolation order parameter ($P_{max}$) for different system sizes ($L=100, 200, 400, 500$) are plotted for pure Ising model; crossing point determines the critical point ($h_p^c=2.10\pm0.01$). The critical Binder cumulant value ($U^*=0.52\pm0.01$) specifies the universality class. Inset shows the data collapse for same $1/\nu$ as obtained before ($0.85$).}
\label{binder}
\end{figure}

To check if the exponents vary with temperature or pulse width, we have scanned the parameter space ($\Delta t, T$) in the range $T=0.5\to 2.0$ and $\Delta t=4\to 10$ with $T=0.5,1.5,2$ and $\Delta t=4,6,10$ (all combinations) apart from the combination (1.0,4) the details of which are shown above. We find no significant variation in the estimates of the critical exponents.  

\begin{figure}[tbh]
\begin{center}
 \centering
 \includegraphics[width=0.9\linewidth]{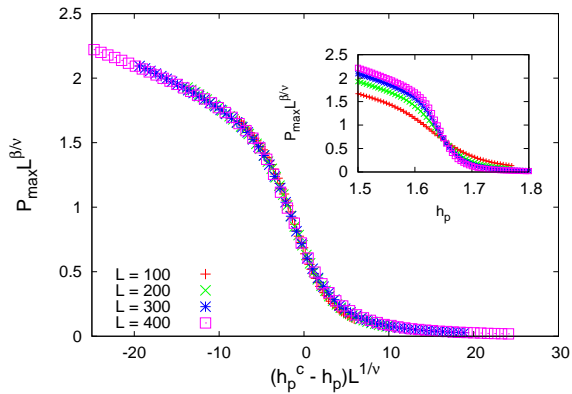}
\end{center}
   \caption{(Color online) Inset: $P_{max}L^{\beta/\nu}$ is plotted against 
field amplitude ($h_p$) for the Ising model with frustration, keeping 
$T=1.0$ and
$\Delta t=4$. The curves for different system sizes ($L=100,200,300,400$) cross at $h_p^c=1.65\pm 0.01$ giving the critical point for $\beta/\nu=0.20\pm0.05$. Front: The value obtained for $\beta/\nu$ is used in the y-axis and by tuning the value of $1/\nu$ all the
curves for different system sizes were made to collapse on a single curve, thereby estimating $1/\nu=0.85\pm0.05$. Error bars are smaller than the system size.}
\label{new-nn-scaleins}
\end{figure}

 The crossing point of the different curves for different system sizes gives the critical point ($2.102\pm 0.002$), which is in good agreement with the previous estimation from finite size scaling ($2.105\pm 0.005$). The value of $U$ at the critical point in this case is $0.52\pm 0.01$ (see Fig.~\ref{binder}). This value remains unchanged for different sets of (mentioned above) temperatures and field pulse width. In the inset of Fig.~\ref{binder} the data collapse is shown. The same value of the exponent $\nu$ is used, as was obtained before.

Therefore, with the increase of temperature and/or field pulse width we find no crossover behavior  in the percolation transition as was reported in \cite{chat} with the magnetisation transition (occurring within the range of the parameter space scanned here). There is indeed a single regime as far as the critical exponents and Binder cumulant values of the percolation transition are considered. 

\begin{figure}[tbh]
\begin{center}
 \centering
 \includegraphics[width=0.9\linewidth]{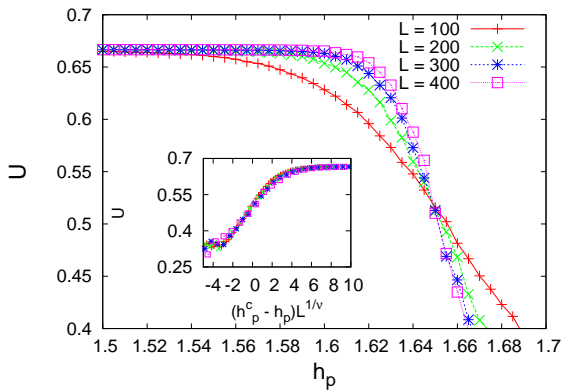}
\end{center}
   \caption{(Color online) Fourth order reduced Binder cumulant of the percolation order parameter ($P_{max}$) for different system sizes ($L=100, 200, 300, 400$) is plotted for the Ising model with frustration; crossing point determines the critical point ($h_p^c=1.65\pm0.01$). The critical Binder cumulant value ($U^*=0.52\pm0.01$) specifies the universality class. Inset shows the data collapse for same $1/\nu$ as obtained before ($0.85$).}
\label{new-binder}
\end{figure}

\subsection{Ising model with second neighbor frustration}
\noindent The critical exponents of the dynamical percolation transition in this case also are found using the finite size scaling analysis as is shown in the pure Ising case. The exponents $\beta/\nu=0.20\pm 0.05$ and $1/\nu=0.85\pm 0.05$ are found from data collapse shown in 
Fig.~(\ref{new-nn-scaleins}).


 The values compare very well with the estimates from the pure Ising model. Fig. (\ref{new-binder}) shows the behavior of the Binder cumulant of the percolation order parameter near the critical point and also its finite size scaling using eqn (\ref{binder-nu}). The critical Binder Cumulant value ($0.52\pm 0.01$) shows the expected agreement with the pure Ising case, as well as the previously estimated value of $1/\nu$ is confirmed from the data collapse shown in the inset.

\section{Discussion and Conclusion}
\noindent It is well studied that for pure Ising model in two dimension, a phase transition in magnetisation and a 
percolation transition of the geometric clusters of nearest neighbor parallel spins occur at the same critical point. 
There have been successful attempts \cite{fort2,fort} to show that this percolation transition is universal in the sense that its
exponents are same for models in the same universality class, providing a new set of exponents that are characteristic features
of Ising and other universality classes in terms of percolation.

It is well known that pure Ising model can undergo a dynamical phase transition (magnetisation reversal) when subjected to
competing pulsed magnetic field. This dynamical phase transition was claimed to belong to a different universality class than
the static transition. However, we find that the critical Binder cumulant value reported in Ref~\cite{chat} in not universal 
for that transition. It changes if the transition is studied, for example, in a different temperature or pulse duration etc.

Here we have studied the same transition in terms of the percolation properties of the geometric clusters.The dynamical transition exponents  were found using finite size scaling analysis. The values of these exponents reported here ($\beta/\nu=0.20\pm 0.05$ and $\nu=1.2 \pm 0.1$), differ significantly from those ($\beta_s/\nu_s=0.052\pm0.002$ and $\nu_s=1.004\pm0.009$) obtained in Ref.~\cite{fort2} for the case of the static transition. Therefore, the dynamical percolation transition is in a different universality class than the static one.  

Further, the critical Binder cumulant value which was found to be non-universal for magnetisation (crossover behavior was reported), is found to be universal (doesn't changes with temperature and/or field pulse width, and microscopic details of the models in the same universality class) 
when studied in terms of the percolation order parameter (within a reasonable range of the parameter space, see sec IIIB), which is the expected behavior (having unique value for a given
universality class) of the critical Binder cumulant.

Also, to see if the exponents reported here indeed characterise the dynamical transition in Ising universality class, we have
repeated the same analysis for another model (diagonal next nearest neighbor frustration) 
belonging to the Ising universality class. It is found that the exponents obtained in this model matches very well 
with those of the pure Ising model. The critical Binder cumulant value also takes the expected universal value. 
Therefore it is concluded that the dynamical percolation behavior reported here is also a characteristic feature of 
any model belonging to the Ising universality class.

Finally, we would like to point out that the values of the critical exponents obtained in our study has remarkable agreement with those obtained in \cite{tsakiris} in which the authors study percolation properties of randomly distributed growing clusters using finite size scaling analysis. 

\begin{acknowledgements}
The authors acknowledge many fruitful discussions and suggestions of Prof. B. K. Chakrabarti and Dr. A. Chatterjee. The computational facilities of CAMCS of SINP were used in producing the numerical results.
\end{acknowledgements} 
\bibliography{article}

\end{document}